\documentstyle[prd,aps,preprint,tighten,floats]{revtex}
\begin{document}
\preprint{IASSNS-HEP-96/123,UPR-728-T,hep-th/9612229}
\date{December 1996}
\title{Rotating Intersecting M-Branes}
\author{Mirjam Cveti\v c$^1$
\thanks{E-mail address:cvetic@cvetic.hep.upenn.edu}
and Donam Youm$^2$
\thanks{E-mail address: youm@sns.ias.edu}}
\address{$^1$ Department of Physics  and Astronomy\\ 
University of Pennsylvania, Philadelphia, PA 19104-6396\\and\\
$^2$ School of Natural Science, Institute for Advanced
Study\\ Olden Lane, Princeton, NJ 08540}
\maketitle
\begin{abstract}
{We present intersecting p-brane solutions of eleven-dimensional 
supergravity ($M$-branes) which upon toroidal compactification reduce 
to non-extreme {\it rotating} black holes.  
We identify harmonic functions, associated with each $M$-brane, and 
non-extremality functions, specifying a deviation from the BPS limit.  
These functions are modified due to the angular momentum parameters,  
which specify the rotation along the transverse directions of the 
$M$-branes.   We spell out the intersection rules for the 
eleven-dimensional space-time metric for intersecting (up to three) 
rotating $M$-brane configurations (and a boost along the common 
intersecting direction).}
\end{abstract} 

\section{Introduction}

Intersecting, BPS-saturated $p$-brane configurations in string theory and 
eleven-dimensional supergravity became subjects of intense research since  
such configurations reduce, upon (toroidal) dimensional compactification, to  
BPS saturated black holes with nonsingular horizons (with non-zero
Bekenstein-Hawking (BH) entropy).  
Within the framework of eleven-dimensional supergravity,  
such configurations correspond to the intersecting two-branes and five-branes 
of eleven-dimensional $M$-theory \cite{DVV,KTT} and they are of 
special interest since the $D=10$ backgrounds with NS-NS and R-R 
charges appear on an equal footing when viewed from eleven dimensions.
This approach may shed light on the structure of non-extreme black holes 
from the point of view of $M$-theory, and, in particular, clarify 
the origin of their BH entropy.

A discussion of intersections of certain BPS-saturated $M$-branes along 
with a proposal for intersection rules was
first given in \cite{PT}.  A generalization to a number of 
different harmonic functions specifying intersecting
BPS-saturated $M$-branes, which led to a better understanding
of these solutions and a construction of new intersecting p-brane 
solutions in $D \leq 11$, was presented in  \cite{TM} (see
also related works in Refs. \cite{KTT,DYA,DYB,CS,COS,GKT,BE,BL,PO}).
Specific configurations of that type reduce to the BPS-saturated black 
holes with regular horizons in $D=5$ \cite{TM} and $D=4$ \cite{KTT},  
whose properties are determined by three and four charges (associated with 
the corresponding  harmonic functions), respectively. 
The generalization to intersecting {\it non-extreme} $M$-brane configurations, 
which upon dimensional reduction become {\it non-extreme} static black 
holes, was given in \cite{CT} (see also \cite{DLP}).

The purpose of the present paper is to spell out the structure of the 
intersecting non-extreme {\it rotating} $M$-brane configurations,  
whose rotational parameters are associated with the transverse directions. 
Upon toroidal compactifications, these solutions become the known non-extreme
rotating black holes (of the toroidally compactified Type II theory).
In particular, we identify the relevant functions that specify the 
intersecting $M$-brane configurations and concentrate on the role of the 
rotational parameters.  
The structure of such configurations may shed light on the 
role of angular momenta in the structure of non-extreme rotating black 
holes from the point of view of $M$-theory, and it may shed light on 
the origin of their contribution to the BH entropy.

In the following, we shall discuss the procedure to construct the 
non-extreme {\it rotating} $M$-brane solutions.  We further identify the 
intersecting configurations of two membrane(s)/five-brane(s) (and also include
a boost along the common intersecting direction), and then further
generalize the procedure to intersecting three and four rotating $M$-branes. 
The resulting non-extreme rotating solutions have the 
eleven-dimensional metric and the  field strength of the three-form field 
specified by a number of independent parameters: 
the ``non-extremality'' parameter, representing the deviation
from the BPS-saturated limit, the ``boost'' parameters, specifying  
charges carried by the configuration, and the angular momentum parameters, 
specifying the rotational degrees of freedom in the transverse directions. 
The important information on the eleven-dimensional metric associated with
each brane configuration is now encoded in the {\it modified} harmonic 
functions, which are specified not only by the charges but also by the 
rotational parameters.
We also concentrate on the structure of such configurations when the
non-extremality parameter is taken to approach zero, i.e., when the
configurations correspond to the BPS-saturated limit.

The paper is structured in the following way.  In Section II, we obtain 
rotating, single $M$-brane solutions by uplifting the rotating solutions 
in lower dimensions.  Based upon the observation made from the structure 
of the rotating single $M$-brane solutions, we infer the 
general algorithm for constructing the rotating intersecting $M$-brane 
solutions in Section III.  In section IV, we check the general algorithm 
against the explicit rotating, intersecting $M$-brane solutions which 
are obtained by uplifting multicharged rotating solutions in lower 
dimensions.  In Section V, we discuss the BPS limits of the explicit 
(intersecting) $M$-brane solutions 
derived in the previous sections.   The conclusions are given in Section VI.

\section{Single Rotating $M$-brane Configuration}

Before we discuss explicit forms of the intersecting $M$-brane solutions, 
we fix our notations for the space-time coordinates.  For an intersecting 
$M$-brane solution, the eleven-dimensional space-time coordinates are divided 
into the ``overall world-volume coordinates'' $(t,y_1,...,y_{p-r})$, 
which are (time+internal)-coordinates common to all the constituent 
$M$-branes, the ``relative transverse coordinates'' $(y_{p-r+1},...,y_p)$, 
which are internal to some of the  $M$-branes, and the 
``overall transverse coordinates'' $(x_1,...,x_{10-p})$, which are transverse 
to all of the $M$-branes.   Then, the rotating intersecting $M$-brane 
solutions depend on the eleven-dimensional space-time coordinates only 
through $(r,\theta,\psi_i)$, where $r\equiv(x^2_1+\cdots+x^2_{10-p})^{1/2}$ 
and $(\theta,\psi_i)$ are the rests of the angular-coordinates of the 
transverse space other than the ones $\phi_i$ associated with the 
rotational planes (see below for definitions).  

While non-extreme static solutions \cite{DLP,CT}  
are specified by the harmonic functions (associated with the 
electric/magnetic charges) and the non-extremality function 
(associated with the mass $m$ of 
the Schwarzschild solution) that depend on the charges and the 
non-extremality parameter, the rotating non-extreme solutions
are specified in addition by non-zero angular momentum parameters $l_i$, 
which enter the solution in terms of the following form of harmonic 
functions $g_i$ ($i=1,...,[{{D-1}\over 2}]$):
\begin{equation}
g_i\equiv 1+{{l^2_i}\over{r^2}},
\label{rotli}
\end{equation} 
which appear in the metric also in combinations:
\begin{equation}
{\cal G}_D\equiv\left\{\matrix{\alpha^2+\sum_i^{{D-2}\over 2}\mu^2_ig^{-1}_i 
& {\rm for\ even\ } D\cr
\sum_i^{{D-1}\over 2}\mu^2_i g^{-1}_i
& {\rm for\ odd\ } D}\right .\ ,
\label{rotG}
\end{equation}
and
\begin{equation}
f_D^{-1}={\cal G}_D\prod^{[{{D-1}\over 2}]}_{i=1}g_i , 
\label{rotfD}
\end{equation}
where $\alpha$ and $\mu_i$ are defined as:
\begin{itemize}
\item Even dimensions $D$:
\begin{eqnarray}
\mu_1 &\equiv&\sin\theta,\ \   \mu_2\equiv\cos\theta\sin\psi_1,
\ \  \cdots ,\ \ 
\mu_{{D-2}\over 2}\equiv\cos\theta\cos\psi_1\cdots
\cos\psi_{{D-6}\over 2}\sin\psi_{{D-4}\over 2},\cr
\alpha&\equiv&\cos\theta\cos\psi_1\cdots\cos\psi_{{D-4}\over 2}, 
\label{edef2}
\end{eqnarray}
\item Odd dimensions $D$:
\begin{eqnarray}
\mu_1&\equiv&\sin\theta,\ \  \mu_2\equiv\cos\theta\sin\psi_1,
\ \   \cdots , \ \ 
\mu_{{D-3}\over 2}\equiv\cos\theta\cos\psi_1\cdots
\cos\psi_{{D-7}\over 2}\sin\psi_{{D-5}\over 2},\cr
\mu_{{D-1}\over 2}&\equiv&\cos\theta\cos\psi_1\cdots
\cos\psi_{{D-5}\over 2}.
\label{odef2}
\end{eqnarray}
\end{itemize}

Here, $D-1$ specifies the number of spatial transverse directions associated
with the rotational degrees of freedom.  Note, that when the corresponding 
11-dimensional solutions are compactified down to $D$ space-time
dimensions the role of parameters $l_i$ is that of the angular momentum 
parameters of the Kerr-Neumann black hole solutions. The function  $f_D$ 
also modifies  the harmonic functions associated with the charge sources 
of the static $M$-brane solutions.  When the rotational
parameters $l_i$ become zero, $g_i$ and ${\cal G}_D$ reduce to one and, 
thereby, the modified harmonic functions and the non-extremality functions 
of the rotating (intersecting) $M$-branes reduce to those of static 
(intersecting) $M$-branes.  

Non-extreme rotating membrane and five-brane solutions of the 
eleven-dimensional supergravity can be obtained by uplifting 
the nine ($D=9$) and six 
($D=6$) dimensional charged, rotating black hole solutions, whose charges 
arise from one two-form $U(1)$ gauge field of the NS-NS sector of toroidally
compactified Type II string.  These solutions were constructed (in terms of 
fields of the toroidally compactified heterotic string) in Refs.
\cite{CYfiv,CYrot}.  
Following the standard procedure discussed in Ref.\cite{SCHM}, 
we obtain the following form of ``rotating'' two- and five-brane solutions.
  
The eleven-dimensional metric of the rotating membrane solution can be cast
in the following form:
\begin{eqnarray}
ds_{11}^2&=&T^{-1/3}\left[T(-fdt^2+dy^2_1+dy^2_2)+f^{\prime\,-1}dr^2\right.
\cr
&+&{r^2}\left\{\left(1+{{l^2_1\cos^2\theta}\over{r^2}}+{{(l^2_2\sin^2\psi_1
+l^2_3\cos^2\psi_1\sin^2\psi_2+l^2_4\cos^2\psi_1
\cos^2\psi_2)\sin^2\theta}\over{r^2}}\right)d\theta^2\right.
\cr
&+&\left(1+{{l^2_2\cos^2\psi_1}\over{r^2}}+{{l^2_3\sin^2\psi_1\sin^2\psi_2}
\over{r^2}}\right)\cos^2\theta d\psi^2_1
+\left(1+{{l^2_3\cos^2\psi_2}\over{r^2}}\right)\cos^2\theta
\cos^2\psi_1d\psi^2_2
\cr
&-&2{{l^2_2-l^2_3\sin^2\psi_2-l^2_4\cos^2\psi_2}\over{r^2}}
\cos^2\theta\cos\psi_1\sin\psi_1\cos\psi_2\sin\psi_2d\psi_1d\psi_2
\cr
&-&2{{l^2_2-l^2_3\sin^2\psi_2-l^2_4\cos^2\psi_2}\over{r^2}}\cos\theta
\sin\theta\cos\psi_1\sin\psi_1d\theta d\psi_1
\cr
&-&2{{l^2_3}\over{r^2}}\cos\theta
\sin\theta\cos^2\psi_1\cos\psi_2\sin\psi_2d\theta d\psi_2
\cr
&-&{{4ml_i\mu^2_i\cosh\delta_e}\over{r^6}}f_DTdtd\phi_i
+{{4ml_il_j\mu^2_i\mu^2_j}\over{r^6}}f_DT
d\phi_id\phi_j
\cr
&+&\left.\left.\mu^2_i
\left(g_i+f_DT{{2ml^2_i\mu^2_i}\over{r^8}}\right)
d\phi^2_i\right\}\right]. 
\label{2bran}
\end{eqnarray}
where the modified ``harmonic function'' $T^{-1}$ associated with 
the electric charge $Q\sim 2m\cosh\delta_e\sinh\delta_e$ source is of the 
form:  
\begin{equation}
T^{-1}\equiv 1+f_D
\left({{2m\sinh^2\delta_e}\over{r^{D-3}}}
\right),  
\label{2harm}
\end{equation} 
and the non-extremality functions $f$ and $f^{\prime}$, appearing respectively 
in front of $dt^2$ and $dr^2$ terms, are of the form:
\begin{equation}
f\equiv 1-f_D
\left({{2m}\over{r^{D-3}}}\right), 
\label{f}
\end{equation}
\begin{equation}
f^{\prime}\equiv {\cal G}^{-1}_D-f_D\left({{2m}\over{r^{D-3}}}\right).
\label{fp}
\end{equation}
Note that in the BPS-saturated limit, i.e., the limit in which 
$m\rightarrow 0$ with $me^{2\delta_e}$ a finite constant, these 
non-extremality functions reduce respectively to $f=1$ and $f^{\prime}=
{\cal G}^{-1}_D$.
The non-zero components of the three-form field are of the following form:
\begin{equation}
B^{(11)}_{t12}={{2m\cosh\delta_e\sinh\delta_e}\over{r^{D-3}}}f_DT, 
\ \ \ \ \ \ 
B^{(11)}_{\phi_i12}=-{{2ml_i\mu^2_i\sinh\delta_e}\over{r^{D-3}}}f_DT.
\label{3eform}
\end{equation}
Here, in the above the number of the transverse spatial  
dimensions $D-1$ is 8 and, therefore, $i=1,...,4$.  
The ADM mass density $M_{ADM}/V_2$ and the angular momentum densities 
$J_i/V_2$ per unit $M\,2$-brane volume, and the electric charge $Q$ are of 
the following form:
\begin{equation}
{{M_{ADM}}\over{V_2}}={{3m\Omega_7}\over{8\pi G_N}}(2\cosh^2\delta_e+1), \ \ \ 
{{J_i}\over{V_2}}={{\Omega_7}\over{4\pi G_N}}ml_i\cosh\delta_e, \ \ \ 
Q={{3m\Omega_7}\over{4\pi G_N}}\cosh\delta_e\sinh\delta_e,
\label{phyparm2br}
\end{equation}
where $V_2$ is the volume of the 2-dimensional space internal to the 
$M\,2$-brane, $G_N$ is the eleven-dimensional Newton's constant, 
and $\Omega_n$ denotes the area of the unit n-sphere.  

The following eleven-dimensional metric of the rotating $M$-five-brane 
solution can be obtained in an analogous way by uplifting the magnetically 
charged rotating solution in six-dimensions: 
\begin{eqnarray}
ds_{11}^2&=&F^{-2/3}\left[F(-fdt^2+
dy^2_1+\cdots+dy^2_5)\right.
\cr
&+&f^{\prime\,-1}{dr^2}
+{r^2}\left\{\left(1+{{l^2_1\cos^2\theta}\over{r^2}}+{{l^2_2\sin^2\theta
\sin^2\psi}\over{r^2}}\right)d\theta^2\right.
\cr
&+&\left(1+{{l^2_2\cos^2\psi}\over{r^2}}\right)
\cos^2\theta d\psi^2-2{{l^2_2}\over{r^2}}\cos\theta\sin\theta
\cos\psi\sin\psi d\theta d\psi
\cr
&-&{{4m\cosh\delta_m}\over{r^3}}f_DF(l_1\sin^2\theta 
dtd\phi_1+l_2\cos^2\theta dtd\phi_2)
\cr
&+&{{4ml_1l_2\cos^2\theta\sin^2\theta\sin^2\psi}\over{r^3}}
f_DFd\phi_1d\phi_2
\cr
&+&\sin^2\theta\left(g_1+{{2ml^2_1\sin^2\theta}
\over{r^5}}f_DF\right)d\phi^2_1
\cr
&+&\left.\left.\cos^2\theta\sin^2\psi
\left(g_2+{{2ml^2_2\cos^2\theta\sin^2\psi}
\over{r^5}}f_DF\right)d\phi^2_2
\right\}\right].
\label{fbra}
\end{eqnarray}
Here, the modified harmonic function $F^{-1}$ associated with the magnetic 
charge $P\sim 2m\cosh\delta_m\sinh\delta_m$ source is of the form: 
\begin{equation}
F^{-1}\equiv 1+ f_D
\left({{2m\sinh^2\delta_m}\over{r^{D-3}}}
\right),
\label{harm5br}
\end{equation}
where $D=6$ (therefore, $f^{-1}_D=g_1\cos\theta+g_2\sin\theta$ 
in Eqs. (\ref{fbra}) and (\ref{harm5br})) and the modified non-extremality 
functions are defined in Eqs. (\ref{f}) and (\ref{fp}) with 
$D=6$.
The non-zero components of the three-form field are of the following
form:
\begin{eqnarray}
B^{(11)}_{\phi_1\phi_2\psi}&=&-2mg_1f_D\cosh\delta_m\sinh\delta_m\cos^2\theta, 
\cr
B^{(11)}_{t\phi_1\psi}&=&{{2ml_2f_D\sinh\delta_m\sin^2\theta}\over
{r^2}}, \ \ \ \ 
B^{(11)}_{t\phi_2\psi}={{2ml_1f_D\sinh\delta_m\cos^2\theta}\over
{r^2}}. 
\label{fivten}
\end{eqnarray}
The ADM mass density $M_{ADM}/V_5$ and the angular momentum densities 
$J_i/V_5$ per unit $M\,5$-brane volume, and the magnetic charge $P$ are 
of the following form:
\begin{equation}
{{M_{ADM}}\over{V_5}}={{3m\Omega_4}\over{8\pi G_N}}(\cosh^2\delta_m+2), \ \ \ 
{{J_i}\over{V_5}}={{\Omega_4}\over{4\pi G_N}}ml_i\cosh\delta_m, \ \ \ 
P={{3m\Omega_4}\over{8\pi G_N}}\cosh\delta_m\sinh\delta_m,
\label{phyparm5br}
\end{equation}
where $V_5$ is the volume of the 5-dimensional space internal to the 
$M\,5$-brane.

\section{Intersection Rules}

Based on the structure obtained in the case of a single rotating membrane and
five-brane, we can infer the rules for constructing the intersecting 
rotating $M$-brane solutions.  Such rules can be checked by uplifting the 
corresponding charged, rotating black brane solutions in lower dimensions.  
Analogously to the case of intersecting non-extreme static $M$-brane 
solutions, one can construct the intersecting non-extreme rotating 
$M$-brane solutions by modifying the flat eleven-dimensional metric, 
following the intersection rules (at least for the overall conformal factor 
and the internal part of the metric). 

The eleven-dimensional space-time, internal to the intersecting  rotating 
M-brane configuration, is specified by  a ``harmonic function'' for each 
constituent $M$-brane (associated with the charge source) and the   
``non-extremality functions'' (associated with the Schwarzschild mass), 
which are now modified by the functions ${\cal G}_D$ and $f_D$ (associated 
with the rotational parameters), defined in Eqs. (\ref{rotG}) and 
(\ref{rotfD}).  The space transverse to the $M$-brane reflects the axial 
symmetry that involves the charge sources as well as the rotational 
parameters.  
Note that such a non-extreme rotating configuration {\it does not}  
have an interpretation as an intersection of {\it separate} non-extreme 
rotating $M$-branes, but rather as a {\it bound state} solution with a 
{\it common} non-extremality parameter and {\it common} rotational 
parameters associated with the common transverse spatial directions.

In the following, we shall specify the algorithm which would yield
the intersecting non-extreme rotating $M$-brane solutions
\footnote{An analogous type of construction applies also to intersecting 
p-brane solutions in ten dimensions.}.  
(These intersection rules can be checked for specific cases in which 
the explicit dimensionally reduced solutions correspond to non-extreme 
rotating multi-charged black hole solutions which have been explicitly 
constructed in \cite{CYfiv,CYrot}.)  
We shall summarize the algorithm in the following two steps.  
First, we shall identify the forms of the non-extremality functions and  
the structure of the transverse part due to the non-zero rotational 
parameters.  We shall also identify the modified harmonic functions, 
associated with each $M$-brane source.  As a step two, we shall spell 
out the modifications of the eleven-dimensional space-time due to both 
non-extremality functions and non-zero $M$-brane sources.

{\noindent Step I}  

\begin{itemize}
\item 
First, one makes the following  replacements in the $D$-dimensional 
(transverse) space-time parts of the metric:
\begin{equation}
dt^2  \to   f dt^2 \ , \ \ \ \ \ \ 
dx_n dx_n \to f^{\prime\,-1}dr^2 + r^2d\Omega^2_{D-2}\ , 
\label{ii}
\end{equation}
where  $f$ and $f^{\prime}$ are the non-extremality functions 
defined as:
\begin{equation}
f=1-f_D{{2m}\over{r^{D-3}}}, \ \ \ \ \  
f^{\prime}={\cal G}^{-1}_D-f_D{{2m}
\over{r^{D-3}}}.
\label{modnonex}
\end{equation}
with  $g_i$, $\cal G_D$ and $f_D$  defined in Eqs. (\ref{rotli}),  
(\ref{rotG}) and  (\ref{rotfD}), respectively.
$d\Omega^2_{D-2}$ is the angular parts of the rotating solution in 
$D$ space-time dimensions.  When all the charges of the (constituent) 
$M$-brane(s) are zero, $d\Omega^2_{D-2}$ takes the following form:
\begin{eqnarray}
d\Omega^2_{D-2}&=&\left(1+{{l^2_1\cos^2\theta}\over{r^2}}
+{{K_1\sin^2\theta}\over{r^2}}\right)d\theta^2
\cr
&+&\left(1+{{l^2_{i+1}\cos^2\psi_i}\over{r^2}}+{{K_{i+1}\sin^2\psi_i}
\over{r^2}}\right)\cos^2\theta\cos^2\psi_1\cdots\cos^2\psi_{i-1}d\psi^2_i
\cr
&-&2{{l^2_j-K_j}\over{r^2}}\cos^2\theta\cos^2\psi_1\cdots\cos^2\psi_{i-1}
\cos\psi_i\sin\psi_i\cdots\cos^2\psi_{j-1}\cos\psi_j\sin\psi_jd\psi_id\psi_j
\cr
&-&2{{l^2_{i+1}-K_{i+1}}\over{r^2}}\cos\theta\sin\theta\cos^2\psi_1\cdots
\cos^2\psi_{i-1}\cos\psi_i\sin\psi_id\theta d\psi_i
\cr
&-&f_D{{4ml_i\mu^2_i}\over{r^{D-1}}}dtd\phi_i
+\mu^2_i\left(g_i+f_D{{2ml^2_i
\mu^2_i}\over{r^{D-1}}}\right)d\phi^2_i
\cr
&+&f_D{{4ml_il_j\mu^2_i\mu^2_j}\over
{r^{D-1}}}d\phi_id\phi_j,
\label{neuang}
\end{eqnarray}
where $K_i$ in Eq. (\ref{neuang}) are defined, in general, for a 
$(D-1)$-dimensional transverse space as:
\begin{itemize}
\item  Even dimensions $D$:
\begin{equation}
K_i\equiv l^2_{i+1}\sin^2\psi_i+\cdots+l^2_{{D-2}\over 2}
\cos^2\psi_i\cdots\cos^2\psi_{{D-6}\over 2}\sin^2\psi_{{D-4}\over 2},
\label{edef1}
\end{equation}
\item Odd dimensions $D$:
\begin{eqnarray}
K_i&\equiv&l^2_{i+1}\sin^2\psi_i+\cdots+l^2_{{D-3}\over 2}
\cos^2\psi_i\cdots\cos^2\psi_{{D-7}\over 2}\sin^2\psi_{{D-5}\over 2}
\cr
&+&l^2_{{D-1}\over 2}\cos^2\psi_i\cdots\cos^2\psi_{{D-5}\over 2}, 
\label{odef1}
\end{eqnarray}
\end{itemize}
Here $i,j$ in $\psi_i$ and $\phi_i$ run from 1 to $[{{D-4}\over 2}]$ 
and from 1 to $[{{D-1}\over 2}]$, respectively.  
When charges carried by intersecting $M$-branes are turned on, 
the above expression (\ref{neuang}) for $d\Omega^2_{D-2}$ gets modified 
in the $(\phi_i,\phi_j)$ and $(t,\phi_i)$ components.

Note $d\Omega_{D-2}^2$ in Eq. (\ref{neuang}) corresponds to the angular 
part of the Kerr solution in $D$ space-time dimensions \cite{MP} and 
reduces to the infinitesimal length element of $S^{D-2}$ when all the 
angular momentum parameters $l_i$ are zero.

\item 
Each constituent membrane and five-brane are respectively specified 
by the following form of modified harmonic functions:
\begin{eqnarray}
T^{-1}_n&=&1+f_D{{2m\sinh^2\delta_n}\over r^{D-3}}, \ \ \ \ 
\cr
F^{-1}_n&=&1+f_D{{2m\sinh^2\gamma_n}\over r^{D-3}}.
\label{25harm}
\end{eqnarray}
These harmonic functions modify the overall conformal factor of the
eleven-dimensional metric, as well as each metric component of  the 
space-time internal to the configuration (see below).

\item In the case when the $M$-brane solution has a null isometry,
i.e., intersecting branes have a common string along some  direction $y$, 
one can add a momentum along $y$ by applying the following $SO(1,1)$-boost 
coordinate transformation among the coordinates $t$ and $y$:
\begin{equation}
t^{\prime}=\cosh\delta_{kk}\,t-\sinh\delta_{kk}\,y,  \ \ \ \ 
y^{\prime}=-\sinh\delta_{kk}\,t+\cosh\delta_{kk}\,y   
\label{iou}
\end{equation}
to the non-extreme rotating background.  

Then, the $(t,t)$- and $(y,y)$-components of the metric are modified as
\footnote{This procedure is analogous to the one
used in introducing a boost  along the common intersection of non-extreme 
static $M$-brane configurations \cite{CT}.}
\begin{eqnarray}
-fdt^2+dy^2\to &-&fdt^{\prime\,2}+dy^{\prime\,2}
\cr
&=&-dt^2+dy^2+f_D{{2m}\over{r^{D-3}}}
(\cosh\delta_{kk} dt-\sinh\delta_{kk} dy)^2
\cr
&=&-K^{-1}fdt^2+K\widehat{dy}^2,\ \ \ \ \  
\widehat{dy}\equiv dy+[K^{\prime\,-1}-1]dt, 
\label{kkboost}
\end{eqnarray}
where
\begin{eqnarray}
K&=&1+f_D{{2m\sinh^2\delta_{kk}}\over{r^{D-3}}},
\cr 
K^{\prime\,-1}&=&1-f_D{{2m\sinh^2\delta_{kk}}\over{r^{D-3}}}K^{-1}.
\label{kkharm}
\end{eqnarray}

In addition, $(\widehat{y_a},\phi_i)$ components (corresponding to the 
induced magnetic field due to non-zero angular momenta) of the 
metric are induced and $(t,\phi_i)$ components get modified. 
\end{itemize}

{\noindent Step II}

\begin{itemize}
\item 
The modified harmonic functions (\ref{25harm}) rescale the the overall 
conformal factor of the eleven-dimensional metric by factors 
$T^{-1/3}_n$ and $F^{-2/3}_m$, for the membrane and five-brane
sources
\footnote{This  structure is the same as in the case of static intersecting 
BPS-saturated \cite{PT,TM} and non-extreme \cite{DLP,CT} M-branes.},  
respectively, and therefore the conformal factor of the intersecting 
$M$-brane solution has the form:
\begin{equation}
ds^2_{11}=(\prod_nT_n)^{-1/3}(\prod_mF_m)^{-2/3}[\cdots],
\label{conf}
\end{equation}
where the terms in the brace $[\cdots]$ are the rest of the metric that 
has to be constructed following along the lines spelled out bellow. 
\item 
The $(t,t)$-component in $[\cdots]$ of Eq. (\ref{conf}) is modified in 
the following way:
\begin{equation}
f(r,\theta,\psi_i)dt^2 \to (\prod_nT_n)(\prod_mF_m)f(r,\theta,\psi_i)dt^2, 
\label{time}
\end{equation}
whereas the $(r,r)$-component in $[\cdots]$ remains of the form 
$f^{\prime\,-1}(r,\theta,\psi_i)dr^2$.  
\item 
There is a factor of $T_n$ [$F_n$] for the $(y_a,y_a)$-components 
(in $[\cdots]$ of Eq. (\ref{conf})), i.e., the components of the internal 
part of the membrane [five-brane] solution, for each constituent 
membrane [five-brane].  Therefore, the internal components are modified 
as:
\begin{eqnarray}
& &dy_1dy_1+dy_2dy_2 \to T(dy_1dy_1+dy_2dy_2),\ \ \  {\rm for\ a\ single\ 
membrane}, 
\cr
& &dy_1dy_1+\cdots+dy_5dy_5 \to F(dy_1dy_1+\cdots+dy_5dy_5),\ \ \  
{\rm for\ a\ single\ fivebrane},
\label{memfivint}
\end{eqnarray}
and for an internal coordinate $y_c$ where membranes and fivebranes 
with harmonic functions $T_m$ and $F_n$ intersect:
\begin{equation}
dy_cdy_c \to (\prod_mT_m)(\prod_nF_n)dy_cdy_c.
\label{intersec}
\end{equation}
Note that p-branes of the same type can intersect over a $(p-2)$-brane
\footnote{These rules were first described in Ref. \cite{PT} by using 
the space-time dimensionality argument of the $M$-brane solutions 
in Ref. \cite{GUV} and can be checked by explicit embeddings of 
lower dimensional black hole solutions into eleven-dimensions.   
These rules are also compatible with the intersection rules for D-branes 
\cite{DBR} of $D=10$ Type II string theory.}, 
i.e., membranes $(p=2)$ can intersect over a 0-brane 
and five-branes $(p=5)$ can intersect over a three-brane.  A membrane 
and a five-brane intersect over a string.

In addition, one can also incorporate the boost transformation (\ref{iou}) 
along the common string along the lines specified at the end of Step I.

\item
The angular component $d\Omega^2_{D-2}$ in the transverse parts has  
information on rotations of (intersecting) $M$-brane(s).  Prior to the 
knowledge of general construction rules for $d\Omega^2_{D-2}$, the 
expressions for $d\Omega^2_{D-2}$ can be inferred from the angular 
components of the $D$-dimensional charged, rotating black brane solutions 
with the same corresponding charge configurations.  When all the angular 
momentum parameters $l_i$ are zero, $d\Omega^2_{D-2}$ is given by the 
infinitesimal length element of (``flat'') $S^{D-2}$ even when $M$-brane 
charges are non-zero.  However, once the angular momenta are turned on, 
the expression for $d\Omega^2_{D-2}$ is altered due to the non-zero 
$M$-brane source charges and becomes increasingly complicated with the 
increase in the number of $M$-brane sources.  

We were able to find general rules for constructing the transverse 
part of intersecting {\it two} and {\it three} $M$-brane solutions 
with non-zero $[{{D-1}\over 2}]$ angular momentum parameters of the 
$D$-dimensional transverse space-time
\footnote{In some of these cases, the corresponding rotating solutions in 
the compactified lower-dimensional theories are known.  On the other hand,  
we do not have a general algorithm for the transverse part of the  
intersecting {\it four} $M$-branes.}.  
In these cases, the expressions for $d\Omega^2_{D-2}$ are given as follows:

(1) Configurations with intersecting {\it two} $M$-branes:
\begin{eqnarray}
d\Omega^2_{D-2}&=&\left(1+{{l^2_1\cos^2\theta}\over{r^2}}
+{{K_1\sin^2\theta}\over{r^2}}\right)d\theta^2
\cr
&+&\left(1+{{l^2_{i+1}\cos^2\psi_i}\over{r^2}}+{{K_{i+1}\sin^2\psi_i}
\over{r^2}}\right)
\cos^2\theta\cos^2\psi_1\cdots\cos^2\psi_{i-1}d\psi^2_i
\cr
&-&2{{l^2_j-K_j}\over{r^2}}\cos^2\theta\cos^2\psi_1\cdots\cos^2\psi_{i-1}
\cos\psi_i\sin\psi_i\cdots\cos^2\psi_{j-1}\cos\psi_j\sin\psi_jd\psi_id\psi_j
\cr
&-&2{{l^2_{i+1}-K_{i+1}}\over{r^2}}\cos\theta\sin\theta\cos^2\psi_1\cdots
\cos^2\psi_{i-1}\cos\psi_i\sin\psi_id\theta d\psi_i
\cr
&-&{{4ml_i\mu^2_i\cosh\delta_1\cosh\delta_2}\over{r^{D-1}}}
f_DH_1H_2dtd\phi_i
\cr
&+&\mu^2_i\left[g_i+f_DH_1H_2{{2ml^2_i
\mu^2_i}\over{r^{D-1}}}
\left(1-f_D{{2m\sinh^2\delta_1\sinh^2\delta_2}
\over{r^{D-3}}}\right)\right]d\phi^2_i
\cr
&+&{{4ml_il_j\mu^2_i\mu^2_j}\over{r^{D-1}}}
\left(1-f_D{{2m\sinh^2\delta_1\sinh^2\delta_2}
\over{r^{D-3}}}\right)f_DH_1H_2
d\phi_id\phi_j,
\label{2chmbr}
\end{eqnarray}
where  $K_i$'s are defined in Eqs. (\ref{edef1}) and (\ref{odef1}) and the 
modified harmonic functions $H_n$, specifying each constituent 
$M$-brane (with electric charges $Q_n$), are given by
\begin{equation}
H^{-1}_n\equiv 1+f_D
{{2m\sinh^2\delta_n}\over{r^{D-3}}}, \ \ \ (n=1,2).
\label{rotharm}
\end{equation}

(2) Configurations with intersecting {\it three} $M$-branes and 
$D-1=4$ transverse spatial dimensions (i.e., a configuration 
corresponding to three membranes intersecting at a point): 
\begin{eqnarray}
d\Omega^2_3&=&(g_1\cos^2\theta+g_2\sin^2\theta)d\theta^2+
{{4m\cos^2\theta\sin^2\theta}\over{r^4}}
\left[l_1l_2\left(1-f_D{{2m}\over{r^2}}\sum_{n<m}
\sinh^2\delta_n\sinh^2\delta_m\right)\right.
\cr
&+&\left.f_D{{2m}\over{r^2}}\left\{(l^2_1+l^2_2)\prod_i
\cosh\delta_i\sinh\delta_i-2l_1l_2\prod_i\sinh^2\delta_i\right\}
\right]f_DH_1H_2H_3d\phi_1d\phi_2
\cr
&-&{{4m\sin^2\theta}\over{r^4}}\left(l_1\prod_i\cosh\delta_i-f
l_2\prod_i\sinh\delta_i \right)f_DH_1H_2H_3dtd\phi_1
\cr
&-&{{4m\cos^2\theta}\over{r^4}}\left(l_2\prod_i\cosh\delta_i-
fl_1\prod_i\sinh\delta_i \right)f_DH_1H_2H_3dtd\phi_2
\cr
&+&\sin^2\theta\left[f^{-1}_D+\left\{{{l^2_2-l^2_1}\over{r^2}}\left(f^{-1}_D
+\sum^3_n{{2m\sinh^2\delta_n}\over{r^2}}\right)-{{2ml^2_1}\over{r^4}}\right\}
f_DH_1H_2H_3\right.
\cr
& &\left.+{{4m^2}\over{r^6}}\left(l^2_2\sum_{m<n}
\sinh^2\delta_m\sinh^2\delta_n-2l_1l_2\prod^3_n\cosh\delta_n
\sinh\delta_n\right)f^2_DH_1H_2H_3\right]d\phi^2_1
\cr
&+&\cos^2\theta\left[f^{-1}_D+\left\{{{l^2_1-l^2_2}\over{r^2}}
\left(f^{-1}_D+\sum^3_n{{2m\sinh^2\delta_n}\over{r^2}}\right)
-{{2ml^2_2}\over{r^4}}\right\}
f_DH_1H_2H_3\right.
\cr
& &\left.+{{4m^2}\over{r^6}}\left(l^2_1\sum_{m<n}
\sinh^2\delta_m\sinh^2\delta_n-2l_1l_2\prod^3_n\cosh\delta_n
\sinh\delta_n\right)f^2_DH_1H_2H_3\right]d\phi^2_2, 
\label{3chmbr1}
\end{eqnarray}
where the modified harmonic functions associated with the 
charge sources $Q_n$ are given by
\begin{equation}
H^{-1}_n=1+f_D{{2m\sinh^2\delta_n}\over{r^2}}, \ \ \ 
(n=1,2,3)
\label{3harm}
\end{equation}
with $f^{-1}_D=g_1\cos^2\theta+g_2\sin^2\theta$.

\item  For the non-zero components $B^{(11)}_{MNP}$ of the eleven-dimensional 
three-form field ${\cal B}$, which can be obtained by uplifting the 
$D$-dimensional expressions for two-form $U(1)$ fields and the two-form 
field, we were able to obtain a general rule for 
the case of intersecting  membranes.  That is, for 
intersecting membranes (intersecting at a point) each with electric charges 
$Q_n$ and internal coordinates $y_{a_n}$ 
and $y_{b_n}$:
\begin{eqnarray}
B^{(11)}_{ta_{n}b_{n}}&=&{{2m\cosh\delta_n\sinh\delta_n}\over
{r^{D-3}}}f_DT_n, 
\cr 
B^{(11)}_{\phi_ia_nb_n}&=&-{{2ml_i\mu^2_i\sinh\delta_n}\over
{r^{D-3}}}f_DT_n.
\label{3rule1}
\end{eqnarray}
For other types of intersecting $M$-brane configurations discussed 
in the following sections, we have not found  a general pattern for 
the three-form fields, and, therefore, we 
simply refer to the following sections for the expressions 
in these cases.  
\end{itemize}

We can now illustrate the above algorithm from several examples of 
explicit eleven-dimensional solutions uplifted from the known 
lower-dimensional non-extreme rotating black hole solutions.  Specifically, 
the single rotating membrane and five-brane described in the previous 
section are of the form satisfying these rules specific for $D=9$ and $D=6$.

In the following section, we turn to the case of intersecting $M$-branes 
and explicitly check the algorithm by uplifting the corresponding 
multi-charged, non-extreme, rotating solutions 
to the eleven dimensions. 

\section{Intersecting Rotating $M$-branes}
\subsection{Intersection of two $M$-branes}

Prior to the knowledge of the algorithms described in the 
previous section, one can construct the intersecting $M$-brane solutions 
with two charges by uplifting the lower-dimensional black hole 
solutions carrying two charges of two-form $U(1)$ gauge fields associated 
with the NS-NS or RR sector of toroidally compactified Type II string 
\cite{CYfiv,CYrot}.  

The eleven-dimensional metric $G^{(11)}_{MN}$ is related to the 
$D$-dimensional (Einstein-frame) metric $g_{\mu\nu}$ in the 
following way.  First, we construct the ten-dimensional metric 
$\hat{G}^{(10)}_{\hat{\mu}\hat{\nu}}$ following the standard Kaluza-Klein 
Ansatz given by:
\begin{equation}
\hat{G}^{(10)}_{\hat{\mu}\hat{\nu}}=
\left(\matrix{e^{a\varphi}g_{\mu\nu}+G_{mn}A^{(1)\,m}_{\mu}
A^{(1)\,n}_{\nu}& A^{(1)\,m}_{\mu}G_{mn}\cr
A^{(1)\,n}_{\nu}G_{mn}& G_{mn}}\right),
\label{10dmet}
\end{equation}
where $\varphi\equiv\hat{\Phi}-{1\over 2}\ln{\rm det}G_{mn}$ is the 
$D$-dimensional metric (this definition of $\varphi$ is used to 
construct the ten-dimensional dilaton $\hat{\Phi}$ from the $D$-dimensional 
solutions for $\varphi$ and $G_{mn}$) and 
$a\equiv{{2}\over{D-2}}$.  Then, we construct the eleven-dimensional 
space-time metric from the ten-dimensional metric and dilaton 
using the Kaluza-Klein Ansatz corresponding to the compactification 
from the eleven dimensions to the ten dimensions:
\begin{equation}
G^{(11)}_{MN}=\left(\matrix{e^{-{{\hat{\Phi}}\over 3}}G^{(10)}_{\hat{\mu}
\hat{\nu}}& e^{{2\over 3}\hat{\Phi}}B_M\cr e^{{2\over 3}\hat{\Phi}}B_N& 
e^{{2\over 3}\hat{\Phi}}}\right),
\label{11dmet}
\end{equation}
where $B_M$ is the ten-dimensional $U(1)$ field in the RR-sector 
of type-IIA string.  
Also, the ten-dimensional two-form field $B^{(10)}_{\hat{\mu}\hat{\nu}}$ 
in the NS-NS sector and three-form field $B^{(10)}_{\hat{\mu}
\hat{\nu}\hat{\rho}}$ in the RR sector are defined from 
the eleven-dimensional three-form field $B^{(11)}_{MNP}$ as 
$B^{(11)}_{MNP}=(B^{(10)}_{\hat{\mu}\hat{\nu}\hat{\rho}},
B^{(11)}_{\hat{\mu}\hat{\nu}10}\equiv B^{(10)}_{\hat{\mu}\hat{\nu}})$.

In order to obtain all the possible intersecting two $M$-brane solutions, 
we begin with the intersecting $M$-brane solution whose lower-dimensional 
counterpart is explicitly known.  
The only such an intersecting $M$-brane solution available at this point 
is the rotating $M$-brane solution corresponding to intersection of 
membrane and fivebrane intersecting at a string $(2\perp 5)$.  
Such a solution can be obtained by uplifting five-dimensional rotating 
black hole solution ($D=5$) \cite{CYfiv} carrying the electric charge of 
the NS-NS two-form $U(1)$ gauge field and the electric charge associated 
with the NS-NS two-form antisymmetric tensor (or the ``magnetic''
charge of the corresponding uplifted eleven-dimensional three-form field) 
of the toroidally compactified heterotic string \cite{CYfiv}.  
The eleven-dimensional metric is of the form:
\begin{eqnarray}
ds^2&=&F^{-2/3}T^{-1/3}\left[FT(-fdt^2+dy^2_1)+F(dy^2_2+dy^2_3+dy^2_4+
dy^2_5)+Tdy^2_6\right.
\cr 
&+&\left.f^{\prime\,-1}{dr^2}+r^2d\Omega^2_3\right]
\label{fivtwo}
\end{eqnarray}
where the modified harmonic functions corresponding to membrane charge 
source $Q\sim 2m\cosh\delta_e\sinh\delta_e$ and five-brane charge source  
$P\sim 2m\sinh\delta_m\cosh\delta_m$ are respectively given by
\begin{eqnarray}
F^{-1}&\equiv&1+f_D
{{2m\sinh^2\delta_m}\over{r^2}}, 
\cr
T^{-1}&\equiv&1+f_D
{{2m\sinh^2\delta_e}\over{r^2}}, 
\end{eqnarray}
with $f^{-1}_D=g_1\cos^2\theta+g_2\sin^2\theta$. 
Here, the angular component $d\Omega^2_3$ is defined in 
Eq. (\ref{2chmbr}), 
and the non-extremality functions $f$ and $f^{\prime}$ are 
respectively given by Eqs. (\ref{f}) and (\ref{fp}) with $D=5$.  
Non-zero components of the eleven-dimensional three-form field 
are given by
\begin{eqnarray}
B^{(11)}_{t16}&=&{{2m\cosh\delta_e\sinh\delta_e}\over
{r^2}}f_DT,
\cr 
B^{(11)}_{\phi_1\phi_26}&=&-2m\cos^2\theta\cosh\delta_m\sinh\delta_m
\left(g_1+{{2mf_D\sinh^2\delta_e}\over{r^2}}\right)T,
\cr
B^{(11)}_{t\phi_16}&=&{{2ml_2\sin^2\theta\cosh\delta_e\sinh\delta_m}
\over {r^2}}f_DT, \ \ \ 
B^{(11)}_{t\phi_26}={{2ml_1\cos^2\theta\cosh\delta_e\sinh\delta_m}
\over {r^2}}f_DT, 
\cr
B^{(11)}_{\phi_116}&=&-{{2ml_1\sin^2\theta\sinh\delta_e\cosh\delta_m}
\over{r^2}}f_DT, \ \ \ 
B^{(11)}_{\phi_216}=-{{2ml_2\cos^2\theta\sinh\delta_e\cosh\delta_m}
\over{r^2}}f_DT. 
\label{tfthree}
\end{eqnarray}

The ADM mass density $M_{ADM}/V_6$ and the angular momentum densities 
$J_i/V_6$ per unit intersecting $M\,2$-brane and $M\,5$-brane volume, the 
electric charge $Q$, and magnetic charge $P$ are of the following form:
\begin{eqnarray}
{{M_{ADM}}\over{V_6}}&=&{{3m\Omega_3}\over{8\pi G_N}}(2\cosh^2\delta_e+
\cosh^2\delta_m), \ \ \ 
{{J_i}\over{V_6}}={{\Omega_3}\over{4\pi G_N}}ml_i\cosh\delta_e\cosh\delta_m, 
\cr
Q&=&{{3m\Omega_3}\over{4\pi G_N}}\cosh\delta_e\sinh\delta_e,\ \ \ 
P={{3m\Omega_3}\over{8\pi G_N}}\cosh\delta_m\sinh\delta_m,
\label{phyparm25br}
\end{eqnarray}
where $V_6$ is the volume of the 6-dimensional space internal to the 
intersecting $M\,2$-brane $M\,5$-brane.  

The rest of the intersecting two $M$-brane solutions, i.e., 
two membranes intersecting at a point ($2\perp 2$) and two 
five-branes intersecting at a three-brane ($5\perp 5$), can be obtained by 
imposing the $T$-duality transformation of the toroidally compactified 
type-II string on the above intersecting membrane and five-brane solution 
($2\perp 5$) compactified on a torus.  The $T$-duality transformations of the 
toroidally compactified type-II string transform type-IIA [type-IIB] 
(intersecting) $M$-brane solutions into type-IIB [type-IIA] (intersecting) 
$M$-brane solutions, while dualizing the overall world-volume directions 
and the overall transverse directions into each other, and dualizing 
the relative transverse directions internal to one $M$-brane to 
the ones internal to the other $M$-brane.  
For the type-II string with the coordinate $x$ compactified on a circle, 
the relevant parts of the $T$-duality transformations necessary in 
constructing the eleven-dimensional metric of the intersecting two 
$M$-branes are given by
\footnote{The complete $T$-duality transformation of the type-II string 
compactified on a torus can be found in Ref.\cite{TII}.}
\begin{equation}
\hat{G}^{(10)}_{xx}\to \hat{G}^{(10)\,\prime}_{xx}={1\over
{\hat{G}^{(10)}_{xx}}}, \ \ \ \ 
\hat{G}^{(10)}_{\mu\nu}\to \hat{G}^{(10)}_{\mu\nu}\ \ (\mu,\nu \neq x),\ \ \ \ 
e^{2\hat{\Phi}}\to e^{2\hat{\Phi}^{\prime}}={{e^{2\hat{\Phi}}}\over 
{\hat{G}^{(10)\,\prime}_{xx}}}.
\label{tdual}
\end{equation}

In general, only $T$-duality transformations that decrease the 
dimensionality of the overall transverse directions guarantee that 
the $T$-dualized solutions satisfy the Euler-Lagrange equations, since 
the other types of $T$-duality transformations introduce additional 
transverse spatial coordinates that the transformed solutions are required 
to depend on.  But since there is only a limited number of available known 
explicit solution (i.e., only $2\perp 5$ configuration) from which we have 
to construct other types of intersecting two $M$-brane solutions through 
the $T$-duality transformations, we shall also use the types of 
$T$-duality transformations that increase the number of dimensions of 
the overall transverse directions.  Namely, we shall assume that the 
$T$-dualized solutions depend on the new introduced transverse spatial 
coordinate, say $x_{10-p}$, as well, through the radial coordinate 
$r=(x^2_1+\cdots +x^2_{10-p})^{1/2}$ of the transverse space, rather than 
becoming delocalized \cite{DEL}. 

The intersecting two membrane solution ($2\perp 2$) in eleven 
dimensions is obtained as follows.   
One first compactifies the coordinate $y_5$ of the solution (\ref{fivtwo})
on a circle to obtain the type-IIA string solution where membrane and 
four-brane intersect at the coordinate $y_1$.  Then, one imposes $T$-duality 
transformation along the coordinate $y_1$ compactified on a circle, 
resulting in an intersecting $p$-brane solution in type-IIB string 
where the string and the three-brane intersect over a zero-brane.  
One transforms this type-IIB solution into a type-IIA solution by 
imposing a $T$-duality transformation
along the coordinate $y_4$ compactified on a circle, resulting in a two  
membranes intersecting at a point.  Finally, one uplifts this type-IIA 
solution to eleven dimensions to obtain  the intersecting two   
membrane solution.

The explicit form of this rotating $M$-brane solution is then given 
as follows:
\begin{eqnarray}
ds^2&=&T^{-1/3}_1T^{-1/3}_2\left
[-T_1T_2fdt^2+T_1(dy^2_1+dy^2_2)+T_2(dy^2_3+dy^2_4)
\right.
\cr
&+&f^{\prime\,-1}dr^2 +{r^2}\left\{(1+{{l^2_1\cos^2\theta}\over{r^2}}
+{{(l^2_2\sin^2\psi+l^2_3\cos^2\psi)\sin^2\theta}\over{r^2}})
d\theta^2\right.
\cr
&+&(1+{{l^2_2\cos^2\psi}\over{r^2}})\cos^2\theta d\psi^2
-2{{l^2_2}\over{r^2}}\cos\theta\sin\theta\cos\psi\sin\psi d\theta d\psi
\cr
&+&{{4ml_il_j\mu^2_i\mu^2_j}\over{r^4}}
(1-f_D{{2m\sinh^2\delta_1\sinh^2\delta_2}\over{r^4}})
f_DT_1T_2d\phi_id\phi_j
\cr
&-&{{4ml_i\mu^2_i\cosh\delta_1\cosh\delta_2}\over{r^4}}
f_DT_1T_2dtd\phi_i
\cr
&+&\left.\left.\mu^2_i[g_i+fT_1T_2
{{2ml^2_i\mu^2_i}\over{r^6}}(1-f_D
{{2m\sinh^2\delta_1\sinh^2\delta_2}\over{r^4}})]d\phi^2_i
\right\}\right], 
\label{int2}
\end{eqnarray}
where the modified harmonic functions $T_n$ associated with 
electric charges $Q_n$ are given by
\begin{equation}
T^{-1}_n=1+f_D{{2m\sinh^2\delta_n}\over{r^4}}, 
\ \ \ n=1,2
\end{equation}
and $f$ and $f^{\prime}$ are given in Eqs. (\ref{f}) and (\ref{fp}) with 
$D=7$.
The non-zero components of the eleven-dimensional three-form field are 
given by:
\begin{eqnarray}
B^{(11)}_{ty_1y_2}&=&{{2m\cosh\delta_1\sinh\delta_1}\over
{r^4}}f_DT_1, \ \ \ \ 
B^{(11)}_{\phi_iy_1y_2}=-{{2ml_i\mu^2_i\sinh\delta_1}\over
{r^4}}f_DT_1 
\cr
B^{(11)}_{ty_3y_4}&=&{{2m\cosh\delta_2\sinh\delta_2}\over
{r^4}}f_DT_2, \ \ \ \ 
B^{(11)}_{\phi_iy_3y_4}=-{{2ml_i\mu^2_i\sinh\delta_2}\over
{r^4}}f_DT_2.  
\label{2perp2}
\end{eqnarray}
The ADM mass density $M_{ADM}/V_4$ and the angular momentum densities 
$J_i/V_4$ per unit intersecting two $M\,2$-brane volume, and the electric 
charges $Q_n$ ($n=1,2$) are of the following form:
\begin{eqnarray}
{{M_{ADM}}\over{V_4}}&=&{{3m\Omega_5}\over{8\pi G_N}}(2\cosh^2\delta_1+
2\cosh^2\delta_2-1), 
\cr
{{J_i}\over{V_4}}&=&{{\Omega_5}\over{4\pi G_N}}ml_i\cosh\delta_1\cosh\delta_2, 
\cr
Q_n&=&{{3m\Omega_5}\over{4\pi G_N}}\cosh\delta_n\sinh\delta_n,
\label{phyparm22br}
\end{eqnarray}
where $V_4$ is the volume of the 4-dimensional space internal to the 
intersecting two $M\,2$-brane.  

The intersecting $M$-brane solution where two five-branes intersect over 
a three-brane are obtained as follows.  First, one compactifies 
the coordinate $\phi_2$ of solution (\ref{int2}) (with $l_2=0$) on a 
circle to obtain intersecting two membranes.  
Then, one imposes the $T$-duality transformation 
along the coordinate $y_1$ compactified on a circle 
to obtain type-IIB solution of two three-branes intersecting at a string  
(along the coordinate $y_1$).  This solution is transformed into type-IIA 
solution of two four-branes intersecting at a membrane (along the 
coordinates $y_1$ and $y_5$), by applying the $T$-duality transformation 
along the coordinate $y_5$  compactified on a circle.  Finally, one 
uplifts this solution to eleven-dimensions to obtain two five-branes 
intersecting at a three-brane.

The final form of this rotating $M$-brane configuration is given by:
\begin{eqnarray}
ds^2&=&F^{-2/3}_1F^{-2/3}_2\left
[F_1F_2(-fdt^2+dy^2_1+dy^2_2+dy^2_3)+
F_1(dy^2_4+dy^2_5)+F_2(dy^2_6+dy^2_7)\right.
\cr
&+&f^{\prime\,-1}dr^2+r^2\left\{
d\theta^2-{{4ml\sin^2\theta\cosh\delta_1\cosh\delta_2}\over r}
f_DF_1F_2dtd\phi\right.
\cr
&+&\left.\left.\sin^2\theta[g+f_DF_1F_2{{2ml^2\sin^2\theta}
\over{r^3}}(1-f_D{{2m\sinh^2\delta_1\sinh^2\delta_2}\over r})]
d\phi^2\right\}\right], 
\label{int5}
\end{eqnarray}
where the modified harmonic functions specified by magnetic charges 
$P_n$ are given by 
\begin{equation}
F^{-1}_n=1+f_D{{2m\sinh^2\delta_n}\over r}, 
\ \ \ n=1,2, 
\end{equation}  
with $f^{-1}_D=g\cos^2\theta+\sin^2\theta$. 
The non-zero components of the three-form field are given by
\begin{eqnarray}
B^{(11)}_{t67}&=&{{2mlf_D\sinh\delta_1\cos\theta}\over{r^2}}, \ \ \ \ 
B^{(11)}_{\phi 67}=-2mgf_D\cosh\delta_1\sinh\delta_1\cos\theta, 
\cr
B^{(11)}_{t45}&=&{{2mlf_D\sinh\delta_2\cos\theta}\over{r^2}}, \ \ \ \ 
B^{(11)}_{\phi 45}=-2mgf_D\cosh\delta_2\sinh\delta_2\cos\theta. 
\label{55form}
\end{eqnarray}
Here, in the above $g=1+{{l^2}\over{r^2}}$.  
The ADM mass density $M_{ADM}/V_7$ and the angular momentum densities 
$J_i/V_7$ per unit intersecting two $M\,5$-brane volume, and the magnetic 
charges $P_n$ ($n=1,2$) are of the following form:
\begin{eqnarray}
{{M_{ADM}}\over{V_7}}&=&{{3m\Omega_2}\over{8\pi G_N}}(\cosh^2\delta_1+
\cosh^2\delta_2+1), 
\cr
{{J_i}\over{V_7}}&=&{{\Omega_2}\over{4\pi G_N}}ml_i\cosh\delta_1\cosh\delta_2, 
\cr
P_n&=&{{3m\Omega_2}\over{8\pi G_N}}\cosh\delta_n\sinh\delta_n,
\label{phyparm55br}
\end{eqnarray}
where $V_7$ is the volume of the 7-dimensional space internal to the 
intersecting two $M\,5$-branes.

One can check that the above $M$-brane solutions for intersecting {\it two} 
$M$-branes ($2\perp 5$, $2\perp 2$ and $5\perp 5$) 
satisfy the  general intersection rules discussed in the previous section. 

\subsection{Intersection of three $M$-branes}

The case of intersecting three 
$M$-branes can be obtained by uplifting the five-dimensional 
rotating black hole solution constructed in Ref.\cite{CYfiv}. Namely, 
when the five-dimensional three-charged rotating black hole solution 
in Ref.\cite{CYfiv} is uplifted to transverse space-time components 
of the eleven dimensions assuming that all the three charges are originated 
from the eleven-dimensional three-form field 
\footnote{Note, one of these three charges of the solution in 
Ref.\cite{CYfiv} is the electric charge of the Kaluza-Klein $U(1)$ gauge 
field (i.e, the electric charge associated with the $SO(1,1)$ boost 
along the isometry direction).  However, this assumption can be justified 
by the fact that the solution in \cite{CYfiv} is the generating solution for 
the most general solution \cite{CH} and, therefore, can be related to the 
rotating solution with all the three charges associated with the 
eleven-dimensional three-form field through $U$-duality (of the 
five-dimensional type-II string theory), which {\it do not} change the 
Einstein frame space-time.},  
the $(t,t)$-component of the metric has a suggestive form as a product 
of three modified harmonic functions, with each modified harmonic function 
corresponding to each non-zero $M$-brane source charge, in agreement with 
the intersection rules discussed in the previous section.  
Therefore, we shall take the angular component $d\Omega^2_3$ of the 
intersecting three $M$-brane with the transverse spatial dimension 
$D-1=4$ to be the corresponding angular coordinate components of the 
three-charged five-dimensional rotating black hole solution constructed 
in Ref.\cite{CYfiv}.  
 
The $M$-brane solution where three membranes intersect over a zero 
brane  ($2\perp 2\perp 2$) has $D-1=4$ transverse spatial dimensions. 
Such a configuration should have the form following the general 
intersection rules discussed in section III with the angular 
component $d\Omega^2_3$ given by the corresponding angular components 
of the five-dimensional, rotating, three-charged black hole obtained in 
Ref.\cite{CYfiv}.   
The explicit form of the rotating $M$-brane solution where three membranes  
intersect ($2\perp 2\perp 2$) is then given by:
\begin{eqnarray}
ds^2_{11}&=&(T_1T_2T_3)^{-1/3}\left[-T_1T_2T_3fdt^2+T_1(dy^2_1+dy^2_2)
+T_2(dy^2_3+dy^2_4)+T_3(dy^2_5+dy^2_6)\right.
\cr
& &\ \ \ \ \ \ \ \ \ \ \ \ \ \left.f^{\prime}dr^2+r^2d\Omega^2_3\right]
\label{3intbran}
\end{eqnarray}
where the modified harmonic functions corresponding to the 
electric charge $Q_n\sim 2m\cosh\delta_n\sinh\delta_n$ sources are 
given by
\begin{equation}
T^{-1}_n\equiv 1+f_D{{2m\sinh^2\delta_n}
\over {r^2}}, \ \ \ n=1,2,3,
\label{3intharm}
\end{equation}
with $f^{-1}_D=g_1\cos^2\theta+g_2\sin^2\theta$.  
Here, the angular component $d\Omega^2_3$ is given by Eq. (\ref{3chmbr1}). 

The non-zero components of the harmonic functions are of the following form:
\begin{eqnarray}
B^{(11)}_{t12}&=&{{2m\cosh\delta_1\sinh\delta_1}\over{r^2}}f_DT_1, \ \ \ \ \ 
B^{(11)}_{\phi_i12}=-{{2ml_i\mu^2_i\sinh\delta_1}\over{r^2}}f_DT_1, 
\cr
B^{(11)}_{t34}&=&{{2m\cosh\delta_2\sinh\delta_2}\over{r^2}}f_DT_2, \ \ \ \ \ 
B^{(11)}_{\phi_i34}=-{{2ml_i\mu^2_i\sinh\delta_2}\over{r^2}}f_DT_2, 
\cr
B^{(11)}_{t56}&=&{{2m\cosh\delta_3\sinh\delta_3}\over{r^2}}f_DT_3, \ \ \ \ \ 
B^{(11)}_{\phi_i56}=-{{2ml_i\mu^2_i\sinh\delta_3}\over{r^2}}f_DT_3,
\label{3memform}
\end{eqnarray}
where $\mu_1=\sin\theta$ and $\mu_2=\cos\theta$.

The ADM mass density $M_{ADM}/V_6$ and the angular momentum densities 
$J_i/V_6$ per unit intersecting three $M\,2$-brane volume, and the electric 
charges $Q_n$ ($n=1,2,3$) are of the following form:
\begin{eqnarray}
{{M_{ADM}}\over{V_6}}&=&{{3m\Omega_3}\over{8\pi G_N}}(2\cosh^2\delta_1+
2\cosh^2\delta_2+2\cosh^2\delta_3-3), 
\cr
{{J_1}\over{V_6}}&=&{{\Omega_3}\over{4\pi G_N}}m(l_1\prod_i\cosh\delta_i-
l_2\prod_i\sinh\delta_i), \ \ \  
\cr
{{J_2}\over{V_6}}&=&{{\Omega_3}\over{4\pi G_N}}m(l_2\prod_i\cosh\delta_i-
l_1\prod_i\sinh\delta_i),
\cr
Q_n&=&{{3m\Omega_3}\over{4\pi G_N}}\cosh\delta_n\sinh\delta_n,
\label{phyparm32br}
\end{eqnarray}
where $V_6$ is the volume of the 6-dimensional space internal to the 
intersecting three $M\,2$-branes.  

\subsection{Intersection of four $M$-branes}

As for the intersecting four $M$-brane solutions, there is no available
lower dimensional rotating charged solution to which the 
intersecting four $M$-brane configuration  would be directly related 
via dimensional reduction.   The explicit general rotating black hole 
solution with four non-zero charges, obtained in Ref. \cite{CYfou}, 
carries two electric charges of the Kaluza-Klein $U(1)$ and two-form 
$U(1)$ gauge fields and two magnetic charges of the Kaluza-Klein $U(1)$ 
and two-form $U(1)$ gauge fields.  In principle, by applying $U$-duality 
transformations (e.g., $O(5,5)$ transformations of Type II theory in 
six-dimensions \cite{SENU,SV}) on this known solution one should 
be able to relate the transformed solution to the intersecting 
configuration of rotating four $M$-branes (i.e., $2\perp 2\perp 5\perp 5$) 
or the intersecting configuration of rotating three $M$-branes
with a boost along the common intersection direction (i.e., $5\perp 5\perp 5 
+ boost$).

We postpone such a study and, following the intersection rules 
of Section III, we simply present a general structure for the 
intersection of two membranes and two fivebranes $(2\perp 2\perp 5\perp 5)$, 
which becomes, after a dimensional reduction, a
four-dimensional rotating black hole with four-charges.  
Such an intersecting $M$-brane solution has the following structure:  
\begin{eqnarray}
ds^2_{11}&=&(T_1T_2)^{-1/3}(F_1F_2)^{-2/3}\left[
-T_1T_2F_1F_2fdt^2 \right.
\cr
&+&F_1(T_1dy^2_1+T_2dy^2_3)+F_2(T_1dy^2_2+T_2dy^2_4)
+F_1F_2(dy^2_5+dy^2_6+dy^2_7)
\cr
&+&\left.f^{\prime\,-1}dr^2+r^2d\Omega^2_2\right]
\label{4intmbr}
\end{eqnarray}
where the ``modified'' harmonic functions $T_i$ and $F_i$ associated, 
respectively, with the electric charges $Q_i\sim 2m\cosh\delta_{ei}
\sinh\delta_{ei}$ and the magnetic charges $P_i\sim 2m\cosh\delta_{pi}
\sinh\delta_{pi}$, and the non-extremality functions are given by 
\begin{eqnarray}
T^{-1}_i&\equiv&1+f_D{{2m\sinh^2\delta_{ei}}\over r}, \ \ \ \ 
F^{-1}_i\equiv 1+f_D{{2m\sinh^2\delta_{pi}}\over r}, \ \ \ \ \ \ i=1,2,  
\cr
f&\equiv&1-f_D{{2m}\over r}, \ \ \ \ 
f^{\prime}\equiv f_D\left(1+{{l^2}\over{r^2}}-{{2m}\over r}\right), 
\label{4intharm}
\end{eqnarray}
where $f^{-1}_D=1+{{l^2\cos^2\theta}\over{r^2}}$.  
Here, the explicit expression for $d\Omega^2_2$ is not known yet.  

\section{Intersecting $M$-brane Solutions with a Boost}

Prior to the knowledge of general $M$-brane construction rules, 
rotating $M$-brane solutions with a boost along a null isometry direction 
are obtained by uplifting lower-dimensional rotating charged solutions 
carrying Kaluza-Klein electric charge.  Such a Kaluza-Klein electric charge 
can be viewed as a boost along the internal isometry direction (or the 
internal momentum) from the viewpoint of the lower-dimensional black hole
solutions, and when it is uplifted to ten dimensions it corresponds 
to the momentum of the string along the compactified direction.  

For example, the following intersecting five-brane and membrane with 
a boost along the intersection coordinate $y_1$ is obtained by 
uplifting the general class of five-dimensional rotating charged black 
hole solution in heterotic string \cite{CYfiv} to eleven dimensions:
\begin{eqnarray}
ds^2&=&F^{-2/3}T^{-1/3}\left[FT(-K^{-1}fdt^2+K\widehat{dy_1}^2)+
F(dy^2_2+\cdots+dy^2_5)+Tdy^2_6\right.
\cr
&+&{{4m\sin^2\theta}\over{r^2}}(l_1\sinh\delta_{kk}\cosh\delta_e\cosh\delta_m
-l_2\cosh\delta_{kk}\sinh\delta_e\sinh\delta_m)
f_DTFd\phi_1\widehat{dy_1}
\cr
&+&{{4m\cos^2\theta}\over{r^2}}(l_2\sinh\delta_{kk}\cosh\delta_e\cosh\delta_m
-l_1\cosh\delta_{kk}\sinh\delta_e\sinh\delta_m)
f_DTFd\phi_2\widehat{dy_1}
\cr
&+&f^{\prime\,-1}dr^2+r^2\left\{(g_1\cos^2\theta+g_2\sin^2\theta)d\theta^2+
{{4ml_1l_2\cos^2\theta\sin^2\theta}\over{r^4}}\right.
\cr
& &\times(1-f_D{{2m\sinh^2\delta_m\sinh^2\delta_e}
\over{r^2}})f_DFTd\phi_1d\phi_2
\cr
&-&{{4m\sin^2\theta}\over{r^4}}(l_1\cosh\delta_e\cosh
\delta_m\cosh\delta_{kk}-l_2f\sinh\delta_e\sinh\delta_m
\sinh\delta_{kk})K^{-1}FTdtd\phi_1
\cr
&-&{{4m\cos^2\theta}\over{r^2}}(l_2\cosh\delta_e\cosh\delta_m
\cosh\delta_{kk}-l_1f\sinh\delta_e\sinh\delta_m
\sinh\delta_{kk})K^{-1}FTdtd\phi_2
\cr
&+&[g_1+f_DFT{{2ml^2_1\sin^2\theta}\over{r^4}}
(1-f_D{{2m\sinh^2\delta_m\sinh^2\delta_e}\over{r^2}})]
\sin^2\theta d\phi^2_1
\cr
&+&\left.\left.[g_2+f_DFT{{2ml^2_2\cos^2\theta}\over{r^4}}
(1-f_D{{2m\sinh^2\delta_m\sinh^2\delta_e}
\over{r^2}})]\cos^2\theta d\phi^2_2\right\}\right]
\label{25boost}
\end{eqnarray}
where the modified harmonic functions $F$, $T$ and $K$ associated with 
the electric $Q\sim 2m\cosh\delta_e\sinh\delta_e$, magnetic 
$P\sim 2m\cosh\delta_m\sinh\delta_m$ and boost $Q_{kk}\sim 2m\cosh
\delta_{kk}\sinh\delta_{kk}$ charge sources are given by 
\begin{eqnarray}
F^{-1}&\equiv&1+f_D{{2m\sinh^2\delta_m}\over{r^2}},
\cr
T^{-1}&\equiv&1+f_D{{2m\sinh^2\delta_e}\over{r^2}}, 
\cr
K&\equiv&1+f_D{{2m\sinh^2\delta_{kk}}\over{r^2}}, 
\end{eqnarray}
and the modified infinitesimal length element $\widehat{dy_1}$ 
of the intersection coordinate $y_1$ is given by
\begin{equation}
\widehat{dy_1}\equiv dy_1+(K^{\prime -1}-1)dt, \ \ \ \ 
K^{\prime -1}\equiv 1-{{2m\sinh\delta_{kk}\cosh\delta_{kk}}\over{r^2}}
f^{-1}_DK^{-1}.
\end{equation}

The non-zero components of the three-form fields are given by
\begin{eqnarray}
B^{(11)}_{t16}&=&{{2m\cosh\delta_e\sinh\delta_e}\over{r^2}}f_DT, 
\cr 
B^{(11)}_{\phi_1\phi_26}&=&-2m\cos^2\theta\cosh\delta_m\sinh\delta_m
\left(g_1+{{2mf_D\sinh^2\delta_e}\over{r^2}}\right)T, 
\cr
B^{(11)}_{t\phi_16}&=&{{2m\sin^2\theta(l_2\cosh\delta_e\sinh\delta_m
\cosh\delta_{kk}-l_1\sinh\delta_e\cosh\delta_m\sinh\delta_{kk})}\over
{r^2}}f_DT,
\cr
B^{(11)}_{t\phi_26}&=&{{2m\cos^2\theta(l_1\cosh\delta_e\sinh\delta_m
\cosh\delta_{kk}-l_2\sinh\delta_e\cosh\delta_m\sinh\delta_{kk})}\over
{r^2}}f_DT,
\cr
B^{(11)}_{\phi_116}&=&{{2m\sin^2\theta(l_2\cosh\delta_e\sinh\delta_m
\sinh\delta_{kk}-l_1\sinh\delta_e\cosh\delta_m\cosh\delta_{kk})}\over 
{r^2}}f_DT, 
\cr
B^{(11)}_{\phi_216}&=&{{2m\cos^2\theta(l_1\cosh\delta_e\sinh\delta_m
\sinh\delta_{kk}-l_2\sinh\delta_e\cosh\delta_m\cosh\delta_{kk})}\over 
{r^2}}f_DT.
\label{25bstform}
\end{eqnarray}

The ADM mass density $M_{ADM}/V_6$ and the angular momentum densities 
$J_i/V_6$ per unit intersecting $M\,2$-brane and $M\,5$-brane volume, the 
electric charge charge $Q$, magnetic charge $P$, and the Kaluza-Klein 
electric charge $Q_{kk}$ are of the following form:
\begin{eqnarray}
{{M_{ADM}}\over{V_6}}&=&{{3m\Omega_3}\over{8\pi G_N}}(2\cosh^2\delta_e+
\cosh^2\delta_m+3\cosh^2\delta_{kk}-3), 
\cr
{{J_1}\over{V_6}}&=&{{\Omega_3}\over{4\pi G_N}}m(l_1\cosh\delta_e\cosh\delta_m
\cosh\delta_{kk}-l_2\sinh\delta_e\sinh\delta_m\sinh\delta_{kk}), 
\cr
{{J_2}\over{V_6}}&=&{{\Omega_3}\over{4\pi G_N}}m(l_2\cosh\delta_e\cosh\delta_m
\cosh\delta_{kk}-l_1\sinh\delta_e\sinh\delta_m\sinh\delta_{kk}), 
\cr
Q&=&{{3m\Omega_3}\over{4\pi G_N}}\cosh\delta_e\sinh\delta_e,\ \ \ 
P={{3m\Omega_3}\over{8\pi G_N}}\cosh\delta_m\sinh\delta_m,
\cr
Q_{kk}&=&{{9m\Omega_3}\over{8\pi G_N}}\cosh\delta_{kk}\sinh\delta_{kk},
\label{phyparm25kbr}
\end{eqnarray}
where $V_6$ is the volume of the 6-dimensional space internal to the 
intersecting $M\,2$-brane and $M\,5$-brane.  

Note that when $\delta_e=0$  (and, therefore, $T^{-1}=1$) [$\delta_m=0$ 
(and, therefore, $F^{-1}=1$)], the solution (\ref{25boost}) reduces 
to a special case of a rotating five-brane with a boost [a rotating 
membrane with a boost]. 

It would be of interest to find the explicit expression for intersecting 
three fivebranes with a $SO(1,1)$ boost along the common string 
($5\perp 5\perp 5 + boost$).  
Such a solution becomes the four-dimensional rotating black hole 
with four charges, after the compactification, and is  related 
to the four-dimensional rotating solution constructed in Ref.\cite{CYfou} 
through $U$-duality transformations of the toroidally compactified, 
six-dimensional type-II string.  
 
\section{BPS Limits}

In this section, we shall discuss BPS limits 
(i.e., the limit in which the ADM energy of the $M$-brane configuration 
saturates the  BPS bound) of rotating intersecting $M$-brane solutions. 
For the case of static, spherically symmetric solutions, the BPS-saturated 
solutions are a subset of extreme solutions (defined 
as  solutions for which the event horizon and the Cauchy horizon 
coincide), but not all of the extreme solutions correspond to the 
BPS-saturated solutions
\footnote{See Refs.\cite{CYfou,NBPS} for some of examples of non-BPS 
extreme solutions.  Also, Ref.\cite{ORT} discusses systematic criteria 
for determining the BPS nature of extreme solutions.}.  
However, when the angular momenta are non-zero, the BPS limit and the 
extreme limit are not correlated.  For a generic rotating charged 
solution, the extreme limit is reached before the BPS limit is reached.  
Thus, most of BPS-saturated {\it rotating} solutions have a
naked singularity.  

For solutions embedded in supergravity theories, the BPS-saturated 
solutions preserve some of supersymmetries.   
The BPS limit is of special interest since  (in the case of $N\ge 4$) 
supersymmetry protects classical solutions from being modified by 
quantum corrections, and therefore one can trust their classical form. 
In particular, those are solutions which have been studied in detail
in order to shed light on the connection between the thermodynamic and 
statistical (microscopic) origins of their entropy. 

In the following, we shall obtain the explicit forms of the BPS limits 
for intersecting {\it rotating} $M$-brane solutions discussed in the 
previous sections. The BPS limit is achieved by 
taking the boost parameters $\delta$'s to infinity and the non-extremality 
parameter $m$ to zero in such a way that the charges carried by the 
intersecting $M$-branes are kept as finite constants.  On the other hand,  
the rotational parameters $l_i$ are kept finite in general, however, in this 
case they may produce singular BPS solutions upon compactification to lower 
dimensions (see below).

Note that such explicit solutions might shed light on the understanding of 
the microscopic entropy of rotating  BPS-saturated black hole solutions
\footnote{In \cite{GT}, it was proposed that such a naked singularity of 
BPS-saturated rotating black holes can be resolved by a generalized 
dimensional reduction (i.e., the dimensional reduction including 
massive Kaluza-Klein modes as well) of the corresponding higher-dimensional 
solution, which has a milder singularity, thereby making it possible to 
analyze the thermodynamic entropy of such singular solutions.} 
from the point of the  $M$-brane picture.    

\subsection{Singular BPS limit}

For  BPS limits,  with angular momentum parameters kept as non-zero 
constants, most of the charged, rotating black holes exhibit a naked 
singularity\footnote{For the space-time dimensions higher than $D=5$, 
the charged rotating black holes with {\it one} non-zero angular momentum 
have regular horizon in the BPS limit \cite{HSEN}.}.   
This is also the case for the corresponding BPS limits of rotating 
intersecting $M$-branes, constructed in the previous sections.  
Namely, when we take the limit $m\to 0$ with $l_i$ kept as non-zero 
constants, the event horizon (defined as roots $r=r_H$ of the 
equation $G^{(11)\,rr}=0$) disappears.  Thus, a singularity, which the 
rotating, intersecting $M$-brane solutions may have, becomes a naked 
singularity.
Therefore, in order to reach the regular BPS limit (i.e., the limit in which 
regular horizon exists in directions transverse to the $M$-brane), 
one has to take the angular momentum parameters to zero, thus resulting 
(in most cases) in {\it static} charged configurations.  Even though the 
BPS solutions with non-zero rotational parameters, which have a naked 
singularity, may not have an immediate  relevance for the study of their  
thermodynamic properties, we shall present them anyway.  In addition, we 
shall also present {\it non-singular} BPS limits for some of (intersecting) 
{\it rotating} $M$-brane solutions derived in the previous sections.  

The singular BPS limit (i.e. the limit in which the space-time transverse 
to the configuration has a naked singularity) corresponds to 
a limit with $\delta's\to\infty$ and $m\to 0$ in such  way 
that  charges remain as finite constants (i.e. $Q_n\sim {\cal Q}_n
\equiv 2m\sinh \delta_n\cosh\delta_n=2m(\sinh\delta_n)^2$) while the 
angular momentum parameters $l_i$ are kept finite.  In this case, 
the non-extremality functions and the modified harmonic functions take 
the following forms:
\begin{equation}
f\to 1, \ \ \  f^{\prime}\to {\cal G}^{-1}_D, \ \ \ 
H_n\to 1+f_D{{{\cal Q}_n}\over{r^{D-3}}}.
\end{equation}
In the following, we shall first obtain the singular BPS saturated solutions 
of a single membrane and a single five-brane, and then find expressions 
for singular BPS limits of intersecting two $M$-branes.  

\noindent{$\bullet$ single $M$-brane}

The singular BPS limit of a membrane:
\begin{eqnarray}
ds_{11}^2&=&T^{-1/3}\left[T(-dt^2+dy^2_1+dy^2_2)+{\cal G}_Ddr^2\right.
\cr
&+&{r^2}\left\{\left(1+{{l^2_1\cos^2\theta}\over{r^2}}+{{(l^2_2\sin^2\psi_1
+l^2_3\cos^2\psi_1\sin^2\psi_2+l^2_4\cos^2\psi_1
\cos^2\psi_2)\sin^2\theta}\over{r^2}}\right)d\theta^2\right.
\cr
&+&\left(1+{{l^2_2\cos^2\psi_1}\over{r^2}}+{{l^2_3\sin^2\psi_1\sin^2\psi_2}
\over{r^2}}\right)\cos^2\theta d\psi^2_1
+\left(1+{{l^2_3\cos^2\psi_2}\over{r^2}}\right)\cos^2\theta
\cos^2\psi_1d\psi^2_2
\cr
&-&2{{l^2_2-l^2_3\sin^2\psi_2-l^2_4\cos^2\psi_2}\over{r^2}}
\cos^2\theta\cos\psi_1\sin\psi_1\cos\psi_2\sin\psi_2d\psi_1d\psi_2
\cr
&-&2{{l^2_2-l^2_3\sin^2\psi_2-l^2_4\cos^2\psi_2}\over{r^2}}\cos\theta
\sin\theta\cos\psi_1\sin\psi_1d\theta d\psi_1
\cr
&-&\left.\left.2{{l^2_3}\over{r^2}}\cos\theta
\sin\theta\cos^2\psi_1\cos\psi_2\sin\psi_2d\theta d\psi_2
+g_i\mu^2_id\phi^2_i\right\}\right],  
\label{bps2bran}
\end{eqnarray}
where the modified harmonic function $T$ associated with the electric 
charge source $Q$ is of the form:
\begin{equation}
T^{-1}=1+f_D{{\cal Q}\over{r^6}},\ \ \ {\cal Q}\equiv {m\over 2}e^{2\delta_e}.
\end{equation}
The non-zero component of the three-form field is given by
\begin{equation}
B^{(11)}_{t12}=T^{-1}
\label{2formbps}
\end{equation}

The singular BPS limit of a  five-brane:
\begin{eqnarray}
ds_{11}^2&=&F^{-2/3}\left[F(-dt^2+dy^2_1+\cdots+dy^2_5)\right.
\cr
&+&{\cal G}_D{dr^2}
+{r^2}\left\{\left(1+{{l^2_1\cos^2\theta}\over{r^2}}+
{{l^2_2\sin^2\theta\sin^2\psi}\over{r^2}}\right)d\theta^2\right.
\cr
&+&\left(1+{{l^2_2\cos^2\psi}\over{r^2}}\right)
\cos^2\theta d\psi^2-2{{l^2_2}\over{r^2}}\cos\theta\sin\theta
\cos\psi\sin\psi d\theta d\psi
\cr
&+&\left.\left.g_1\sin^2d\phi^2_1
+g_2\cos^2\theta\sin^2\psi d\phi^2_2\right\}\right], 
\label{bpsfbra}
\end{eqnarray}
where the modified harmonic function $F$ associated with the 
magnetic charge source $P$ is given by
\begin{equation}
F^{-1}=1+f_D{{\cal P}\over{r^3}},\ \ \ {\cal P}\equiv{m\over 2}e^{2\delta_m}.
\end{equation}
The non-zero component of the three-form field is given by
\begin{equation}
B^{(11)}_{\phi_1\phi_2\psi}=-{\cal P}g_1f_D\cos^2\theta
\label{5formbps}
\end{equation}

Note, in the BPS limit the angular momenta of single $M$-branes become 
zero (as one can see from the expressions for angular momenta $J_i$ in 
Eqs.(\ref{phyparm2br}) and (\ref{phyparm5br})), while the metric is 
non-trivially modified
\footnote{This may be an example of a violation of classical ``no-hair'' 
theorem.  Another example can be found in Ref.\cite{HORt}.} 
due to non-zero rotational parameters $l_i$ 
since we are keeping $l_i$ in the solutions Eqs.(\ref{2bran}) 
and (\ref{fbra}) as non-zero constants.  

\noindent{$\bullet$ Intersecting two $M$-branes}

The singular BPS limit of an intersecting membrane and five-brane:
\begin{eqnarray}
ds^2&=&F^{-2/3}T^{-1/3}\left[FT(-dt^2+dy^2_1)+F(dy^2_2+dy^2_3+dy^2_4+
dy^2_5)+Tdy^2_6\right.
\cr 
&+&{\cal G}_D{dr^2}+r^2\left\{(g_1\cos^2\theta+g_2\sin^2\theta)d\theta^2 
-{{2{\cal QP}l_1l_2\cos^2\theta\sin^2\theta}\over{r^4}}f^2_D
FTd\phi_1d\phi_2\right.
\cr
&-&{{2\sqrt{{\cal QP}}l_1\sin^2\theta}\over{r^2}}f_DFTdtd\phi_1
-{{2\sqrt{{\cal Q}{\cal P}}l_2\cos^2\theta}\over{r^2}}f_DFTdtd\phi_2
\cr
&+&\left.\left.\left(g_1-{{{\cal QP}l^2_1\sin^2\theta}\over{r^6}}
f^2_DFT\right)\sin^2\theta d\phi^2_1
+\left(g_2-{{{\cal QP}l^2_2\cos^2\theta}\over{r^6}}f^2_DFT\right)
\cos^2\theta d\phi^2_2, 
\right\}\right]
\label{bpsfivtwo}
\end{eqnarray}
where the modified harmonic functions $T$ and $F$ associated with 
the electric $Q$ and magnetic $P$ charge sources are given by
\begin{equation}
T^{-1}=1+f_D{{\cal Q}\over{r^2}}, \ \ \ \ \ 
F^{-1}=1+f_D{{\cal P}\over{r^2}}.
\end{equation}
The non-zero components of the three-form field are of the following form:
\begin{eqnarray}
B^{(11)}_{t16}&=&T^{-1},\ \ \ \ \ 
B^{(11)}_{\phi_1\phi_26}=-{\cal P}\cos^2\theta\left(g_1+{{{\cal Q}f_D}
\over{r^2}}
\right)T, 
\cr
B^{(11)}_{t\phi_16}&=&{{l_2\sqrt{\cal QP}\sin^2\theta}\over{r^2}}f_DT, 
\ \ \ \ \ 
B^{(11)}_{t\phi_26}={{l_1\sqrt{\cal QP}\cos^2\theta}\over{r^2}}f_DT, 
\cr
B^{(11)}_{\phi_116}&=&-{{l_1\sqrt{\cal QP}\sin^2\theta}\over{r^2}}f_DT, 
\ \ \ \ \ 
B^{(11)}_{\phi_216}=-{{l_2\sqrt{\cal QP}\cos^2\theta}\over{r^2}}f_DT. 
\label{25formbps}
\end{eqnarray}

The singular BPS limit of intersecting two membranes:
\begin{eqnarray}
ds^2&=&T^{-1/3}_1T^{-1/3}_2\left[-T_1T_2dt^2
+T_1(dy^2_1+dy^2_2)+T_2(dy^2_3+dy^2_4)
\right.
\cr
&+&{\cal G}_Ddr^2 +{r^2}\left\{\left(1+{{l^2_1\cos^2\theta}\over{r^2}}
+{{(l^2_2\sin^2\psi+l^2_3\cos^2\psi)\sin^2\theta}\over{r^2}}\right)
d\theta^2\right.
\cr
&+&\left(1+{{l^2_2\cos^2\psi}\over{r^2}}\right)\cos^2\theta d\psi^2
-2{{l^2_2}\over{r^2}}\cos\theta\sin\theta\cos\psi\sin\psi d\theta d\psi
\cr
&-&{{2\sqrt{{\cal Q}_1{\cal Q}_2}l_i\mu^2_i}\over{r^4}}f_DT_1T_2dtd\phi_i
+\mu^2_i\left(g_i-{{{\cal Q}_1{\cal Q}_2l^2_i\mu^2_i}\over{r^{10}}}
f^2_DT_1T_2\right)
d\phi^2_i
\cr
&-&\left.\left.{{2{\cal Q}_1{\cal Q}_2l_il_j\mu^2_i\mu^2_j}\over{r^8}}
f^2_DT_1T_2d\phi_id\phi_j\right\}\right],
\label{bpsint2}
\end{eqnarray}
where the modified harmonic functions $T_n$ associated with the 
electric charge sources $Q_n$ are given by
\begin{equation}
T^{-1}_n=1+f_D{{{\cal Q}_n}\over{r^4}},\ \ \ \ (n=1,2).
\end{equation}
The non-zero components of the three-form field are given by
\begin{equation}
B^{(11)}_{t12}=T^{-1}_1, \ \ \ \ \ 
B^{(11)}_{t34}=T^{-1}_2.
\label{22formbps}
\end{equation}

The singular BPS limit of intersecting two five-branes:
\begin{eqnarray}
ds^2&=&F^{-2/3}_1F^{-2/3}_2\left
[F_1F_2(-dt^2+dy^2_1+dy^2_2+dy^2_3)+
F_1(dy^2_4+dy^2_5)+F_2(dy^2_6+dy^2_7)\right.
\cr
&+&{\cal G}_Ddr^2+r^2\left\{
d\theta^2-{{2\sqrt{{\cal P}_1{\cal P}_2}l\sin^2\theta}\over r}
f_DF_1F_2dtd\phi\right.
\cr
&+&\left.\left.\sin^2\theta\left(g-{{{\cal P}_1{\cal P}_2l^2\sin^2\theta}
\over{r^4}}f^2_DF_1F_2\right)d\phi^2\right\}\right], 
\label{bpsint5}
\end{eqnarray}
where the modified harmonic functions $F_n$ associated with the 
magnetic charge sources $P_n$ are given by
\begin{equation}
F^{-1}=1+f_D{{{\cal P}_n}\over r}, \ \ \ (n=1,2).
\end{equation}
The non-zero components of the three-form fields are given by
\begin{equation}
B^{(11)}_{\phi 67}=-{\cal P}_1gf_D\cos\theta, \ \ \ \ \ 
B^{(11)}_{\phi 45}=-{\cal P}_2gf_D\cos\theta.
\label{55formbps}
\end{equation}

\subsection{Regular BPS limit}

Again for generic non-extreme, charged, rotating black holes, when the regular 
BPS limit is reached, the solutions reduce to static 
solutions.  However, for the general five-dimensional 
rotating black holes with three charges, one can reach the regular BPS limit, 
i.e. the BPS limit with a regular horizon,  
with non-zero angular momenta by choosing one linear combination of angular 
momenta to be non-zero (see e.g., \cite{TSEf,CYfiv}). 

For the intersecting $M$-brane solutions with three charges 
(which is compactified to the above general class of five-dimensional black 
hole with regular BPS limit) $Q_k\sim 2m\cosh\delta_k\sinh\delta_k$, 
$k=1,2,3$ (including electric/magnetic charges of the constituent 
$M$-branes, electric charge associated with the $SO(1,1)$ boost along the 
isometry direction of the internal coordinate, etc), and $D-1=4$ transverse 
space dimensions, the regular BPS limit is 
achieved by taking the limit in which $\delta_k\to\infty$ and $m,l_i\to 0$ 
in such a way that the following combinations remain as constants:
\begin{equation}
{\cal Q}_k={1\over 2}me^{2\delta_k}, \ \ \ \ \ 
L_i={{l_i}\over {\sqrt{m}}}, \ \ \ (k=1,2,3;\ i=1,2).  
\label{bpslm}
\end{equation}

In the regular BPS limit, the non-extremality functions and the 
Harmonic functions take the following forms:
\begin{equation}
g_i, {\cal G}_D, f, f^{\prime}\to 1, \ \ \ \ \ 
H_n\to 1+{{{\cal Q}_n}\over{r^{D-3}}}, 
\end{equation}
and $K_i$ defined in Eqs. (\ref{edef1}) and (\ref{odef1}) become zero.

The first example of the intersecting $M$-brane solution with 
$D-1=4$ transverse spatial dimensions which has the regular 
BPS limit with non-zero angular momenta corresponds to 
the intersecting membrane and fivebrane solution with a boost along 
the $y_1$-directions (Eq. (\ref{25boost})).  The explicit form of the 
BPS limit is given as follows: 
\begin{eqnarray}
ds^2&=&F^{-2/3}T^{-1/3}\left[FT(-K^{-1}dt^2+K\widehat{dy_1}^2)+
F(dy^2_2+\cdots+dy^2_5)+Tdy^2_6\right.
\cr
&+&{{J\sin^2\theta}\over{r^2}}TFd\phi_1\widehat{dy_1}
-{{J\cos^2\theta}\over{r^2}}TFd\phi_2\widehat{dy_1}
+dr^2
\cr
&+&r^2\left.\left\{d\theta^2
-{{J\sin^2\theta}\over{r^4}}K^{-1}FTdtd\phi_1
+{{J\cos^2\theta}\over{r^4}}K^{-1}FTdtd\phi_2
+\sin^2\theta d\phi^2_1+\cos^2\theta d\phi^2_2\right\}\right], 
\label{regbps}
\end{eqnarray}
where the harmonic functions $F$, $T$ and $K$ associated with the 
electric $Q$, magnetic $P$ and the boost electric $Q_{kk}$ charges 
are given by
\begin{equation}
F^{-1}=1+{{\cal P}\over{r^2}}, \ \ \ \ \ \ 
T^{-1}=1+{{\cal Q}\over{r^2}}, \ \ \ \ \ \ 
K=1+{{{\cal Q}_{kk}}\over{r^2}}, 
\label{regbpsdef}
\end{equation}
and the modified infinitesimal line element $\widehat{dy_1}$ along 
the boost direction $y_1$ is given by
\begin{equation}
\widehat{dy_1}=dy_1+(K^{-1}-1)dt.
\end{equation}
Here, the angular momenta are defined in terms of the 
constants in Eq. (\ref{bpslm}) as
\begin{equation}
J\equiv J_{\phi_1}=-J_{\phi_2}=(2{\cal QPQ}_{kk})^{1\over 2}(L_1-L_2)
\label{5dang}
\end{equation}
The non-zero components of the three-form field are given by:
\begin{eqnarray}
B^{(11)}_{t16}&=&T^{-1}, \ \ \ \ \ 
B^{(11)}_{\phi_1\phi_26}=-{\cal P}\cos^2\theta, \ \ \ \ \ 
B^{(11)}_{t\phi_16}=-{{J\sin^2\theta}\over{2r^2}}T,
\cr
B^{(11)}_{t\phi_26}&=&{{J\cos^2\theta}\over{r^2}}T,\ \ \ \ \ 
B^{(11)}_{\phi_116}=-{{J\sin^2\theta}\over {2r^2}}T, \ \ \ \ \  
B^{(11)}_{\phi_216}={{J\cos^2\theta}\over{2r^2}}T. 
\label{25bformbps}
\end{eqnarray}

The second example corresponds to the intersecting three membrane ($2\perp 2
\perp 2$) solution (\ref{3intbran}) with the regular BPS limit having the 
following form:
\begin{eqnarray}
ds^2_{11}&=&(T_1T_2T_3)^{-1/3}\left[-T_1T_2T_3dt^2+T_1(dy^2_1+dy^2_2)
+T_2(dy^2_3+dy^2_4)+T_3(dy^2_5+dy^2_6)\right.
\cr
& &\ \ \ \ \ +dr^2+r^2\left\{d\theta^2+{{J^2}\over{2r^6}}T_1T_2T_3\cos^2\theta
\sin^2\theta d\phi_1d\phi_2\right. 
\cr
& &\ \ \ \ \ -{{2J}\over{r^4}}T_1T_2T_3\sin^2\theta dtd\phi_1
+{{2J}\over{r^4}}T_1T_2T_3\cos^2\theta dtd\phi_2 
\cr
& &\ \ \ \ \ \left.\left.+\sin^2\theta\left(1-{{J^2}\over{4r^6}}T_1T_2T_3
\sin^2\theta\right)d\phi^2_1+\cos^2\theta\left(1-{{J^2}\over{4r^6}}
T_1T_2T_3\cos^2\theta\right)d\phi^2_1\right\}\right], 
\label{3membps}
\end{eqnarray}
where the harmonic functions $T_n$ associated with the electric charge 
$Q_n$ are given by
\begin{equation}
T^{-1}_n=1+{{{\cal Q}_n}\over{r^2}}, \ \ \ n=1,2,3,
\label{3brharm}
\end{equation}
and the angular momenta are defined in terms of the constants 
in (\ref{bpslm}) as
\begin{equation}
J_1=-J_2=J=(2{\cal Q}_1{\cal Q}_2{\cal Q}_3)^{1\over 2}(L_1-L_2).
\label{3brang}
\end{equation}
The non-zero components of the three-form field are given as follows:
\begin{equation}
B^{(11)}_{ta_nb_n}=T^{-1}_n, \ \ \ (n=1,2,3)
\label{3memformbps}
\end{equation}
where $(a_n,b_n)=(1,2),(3,4),(5,6)$  for $n=1,2,3$, respectively.

\section{Conclusions}

In this paper, we set out to address a general structure of non-extreme 
{\it rotating} intersecting $M$-brane configurations of eleven-dimensional 
supergravity.  Such configurations should be interpreted as {\it bound state} 
solutions of $M$-branes with a {\it common} non-extremality parameter 
and {\it common} rotational parameters associated with the transverse 
spatial directions of the $M$-brane configuration.

An important result is  {\it a general algorithm for constructing 
the overall conformal factor and the internal components}  of the 
eleven-dimensional metric for such configurations.  We also spelled out
the action of a boost along the  common intersecting direction.
The space-time describing the {\it internal} part of such (intersecting) 
configurations is specified {\it entirely} by ``harmonic functions'' for each 
constituent $M$-brane (associated with each charge source) and the   
``non-extremality functions'' (associated with the Schwarzschild mass), 
which are, in contrast to the static case, modified by functions that 
depend on the the rotational parameters (see Section III).  
This general algorithm (while inferred
from the structure of the solution for a single rotating membrane and 
a single rotating five-brane) can be checked (and confirmed)  
against the explicit  solutions of intersecting $M$-brane solutions, uplifted 
from the known rotating, charged black hole solutions in lower dimensions 
and the $T$-duality related solutions thereof.  

On the other hand, the transverse part of the configuration, which 
reflects the axial symmetry of the solution, involves charge sources 
as well as the  rotational parameters in a more involved manner, and cannot 
be simply written in terms of modified harmonic functions and non-extremality 
functions, only
\footnote{Note that in the case of static solutions the transverse 
part has a uniform structure of the form $f^{-1}(r)dr^2+r^2d\Omega^2_{D-1}$ 
with $d\Omega^2_{D-1}$ given by the infinitesimal length element of the 
unit $(D-2)$-sphere $S^{D-2}$, independently of the number of non-zero 
charges.}.  
Therefore, in contrast to the case of static solutions, the general 
algorithm for constructing the transverse part of the metric  was obtained 
only for intersecting {\it two} $M$-branes and for intersecting {\it three} 
membranes with the transverse spatial dimensions $D-1=4$. Again this structure 
can be checked against the corresponding explicit solutions uplifted from 
known lower-dimensional solutions.

In contrast to the static solutions, the eleven-dimensional 
three-form field  acquires  additional non-zero components due 
to the electric [magnetic] field induced from rotating magnetic [electric] 
charge sources, thus assuming a complicated structure.  The structure of 
the three-form field was obtained only in the case of intersecting {\it two} 
$M$-branes and the intersecting membrane and fivebrane with the $SO(1,1)$ 
boost, when it can be checked against the explicit solutions.

The hope is that structure of such configurations may shed light on the  
role of angular momentum components of the non-extreme rotating black 
holes from the point of view of $M$-theory, and it may  ultimately shed 
light on the origin of their contribution to the  entropy of such a system.  
While the part internal to the configuration is well understood and has a 
clear interpretation in terms of harmonic functions,  as ``sources'' for 
each $M$-brane (and the non-extremality functions), the structure of the 
transverse part of the metric needs further investigations.  In the BPS 
limit, the structure of the transverse part simplifies significantly and 
therefore it may be possible to write the general structure in a tractable 
form.

\acknowledgments
We  would like to thank G. Gauntlett and A. Tseytlin for useful discussions.  
The work is supported by U.S. DOE Grant Nos. DOE-EY-76-02-3071 (M.C.),
DOE DE-FG02-90ER40542 (D.Y.) and the National 
Science Foundation Career Advancement Award No. PHY95-12732 (M.C.).

\end{document}